\documentclass{svmult}

\usepackage{mathptmx}       
\usepackage{helvet}         
\usepackage{courier}        

\usepackage{makeidx}         
\usepackage{multicol}        
\usepackage[bottom]{footmisc}

\usepackage{cite}
\usepackage{enumitem}

\usepackage[pdftex]{hyperref}
\hypersetup{colorlinks=true,
            urlcolor=blue,
            citecolor=blue,
            pdftitle={The NEMESYS Approach to Mobile Network Anomaly Detection and Mitigation},
            pdfauthor={Omer H. Abdelrahman, Erol Gelenbe, Gokce Gorbil, Boris Oklander}}

\makeindex             

\begin{document}

\title*{Mobile Network Anomaly Detection and Mitigation: The NEMESYS Approach}
\titlerunning{Mobile Network Anomaly Detection and Mitigation}
\author{Omer H. Abdelrahman, Erol Gelenbe, G\"{o}k\c{c}e G\"{o}rbil and Boris Oklander}
\authorrunning{O. H. Abdelrahman et al.}
\institute{O. H. Abdelrahman \and E. Gelenbe \and G. G\"{o}rbil \and B. Oklander
\at Department of Electrical \& Electronic Engineering, Imperial College, London SW7 2BT, UK\\ email: o.abd06@imperial.ac.uk}

\maketitle

\abstract{
Mobile malware and mobile network attacks are becoming a significant threat that accompanies the increasing popularity of smart phones and tablets. Thus in this paper we present our research vision that aims to develop a network-based security solution combining analytical modelling, simulation and learning, together with billing and control-plane data, to detect anomalies and attacks, and eliminate or mitigate their effects, as part of the EU FP7 NEMESYS project. These ideas are supplemented with a careful review of the state-of-the-art regarding anomaly detection techniques that mobile network operators may use to protect their infrastructure and secure users against malware.
}

\section{Introduction}

Mobile malware is emerging as a significant threat due to the increasing popularity of smart phones and tablets, which now run fully-fledged operating systems (OSs) on powerful hardware and feature multiple interfaces and sensors. Personal computers (PCs) are no longer the dominant computing platform, and indeed the global shipments of smart phones alone have exceeded those of desktop computers since 2011 \cite{Canalys2012}. Further, with the accelerated adoption of 4G technologies including WiMAX and LTE, cellular devices will become the primary means of broadband Internet access for many users. In fact, while 4G capable devices represent only 0.9\%  of all global mobile connections observed in the Internet during 2012, they already account for 14\% of the mobile data traffic \cite{Cisco2013}. As more and more people move from PCs to handheld devices, cyber criminals are naturally shifting their attention to the mobile environment, and this trend is fuelled by the availability of off-the-shelf malware creation tools \cite{BBC2012} as well as the proliferation of mobile application (shortly known as app) marketplaces, enabling the distribution of malicious apps to potentially millions of users \cite{lookout2013}. Such mobile malware can attempt to snoop and steal saleable information, generate revenue by calling premium rate numbers, or perform other malicious activities.

Despite this growing challenge, operators continue to be reactive rather than proactive towards these security threats \cite{Arbor2012}, and investments in detection and mitigation techniques specific to mobile networks are only implemented when a problem occurs. In this position paper regarding certain research activities of the EU FP7 project NEMESYS \cite{NEMESYS1}, we describe our approach to \emph{the research and development of anomaly detection techniques so that mobile network operators may protect their own infrastructure and defend users against malware}. The techniques that operators may develop and deploy on their networks can constitute value-added services for communities of users, in the same way that banks use profiling to reduce credit card fraud. If deployed at the network rather than mobile devices, such services will also spare mobile device batteries and bandwidth (both computational and communication). In fact, most users are not aware of the growing security risks with mobile devices \cite{AVG2011}, but it is in the interest of operators to ensure that users are well protected since large scale malware infections pose a significant threat to the availability and security of cellular networks. Network level analysis also provides a broad view of malicious activities within an operator's network, is not vulnerable to exploits that allow malware to circumvent client-side security, and can be modified easily without requiring users to install patches and updates.

In the sequel, we first present our vision about how to address this field of research in the context of NEMESYS for a network-based security solution which combines modelling and learning, and uses network measurements as well as control-plane and billing data to detect and evaluate anomalies and attacks. Then we review the recent literature on attacks targeting services and users of mobile networks, and outline the merits and limitations of existing network- and cloud-based anomaly detection techniques. Finally we draw some conclusions.

\section{The NEMESYS Model-based Approach}

The research we are conducting with regard to anomaly detection and mitigation within the NEMESYS project uses a model-based approach that involves representing how the communication system functions at the level of each mobile connection. The model-based approach is motivated by several factors. \emph{First}, the number of mobile users that we need to monitor and deal with in real time is very large. Thus a clear and understandable uniform approach is needed to deal with each individual mobile call, emphasising the similarities and common parameters. Anomalies can then be detected via deviations from normal parameters or behaviours. \emph{Second}, the computational tools that are being developed for anomaly detection and mitigation need to be based on sound principles; mathematical models allow us to evaluate and validate such algorithms based on clear underlying model assumptions, even though the use of these models in various practical situations will include conditions when some of the model assumptions are not satisfied. Thus mathematical models will need to be tested and validated through simulation and practical measurements. \emph{Third}, due to the sheer size of the systems we need to deal with, the mathematical models will have to be decomposable, both in terms of obtaining analytical and numerical solutions, e.g. in terms of product forms \cite{Muntz}, and in terms of distributed processing for reasons of scalability \cite{Aguilar}. Again, the mathematical and decomposable structure also provides a handle for decomposing and distributing the computational tasks.

The focus of model construction is on identifying and modelling the individual steps that a mobile user makes regarding:
\begin{itemize}[leftmargin=*,topsep=0.2em]
	\item Call establishment, including the connection to base stations, access points, and call management,

	\item Monitoring and billing, and the interactions between the mobile operator's resources and the network for monitoring and billing,

	\item Accesses that the call may make to sensitive resources such as web sites for privileged information interrogation,

	\item Call processing or service steps that may require that the mobile user identity itself to the network or external resources, or provide other sensitive information (e.g. personal addresses) at certain operational steps,

	\item The access to web sites that are used for purchasing and billing.
\end{itemize}
Indeed, in order to develop detection capabilities of a practical value it is vital to formulate a unified analytical framework which explicitly describes the main {\em internal resources} of the network architecture, including both the internal  aspects regarding base station, access points and call management and billing, and the {\em sensitive external resources} that the mobile user may access during its call. Since our approach will have to be effective in situations where hundreds of thousands of mobile users may be active in a given network simultaneously, we need to address both:
\begin{itemize}[leftmargin=*,topsep=0.2em]
	\item The case where only a small percentage of mobile users come under attack at a given time, but these attacks are nevertheless of high value to the attacker so that we must be able to detect relatively rare events in a very large ensemble, a little like detecting a small number of hidden explosive devices in a very large and potentially dangerous terrain, also

	\item Situations where attacks affect the signaling system and are significantly disturbing a large fraction of the ongoing mobile connections.
\end{itemize}
In all cases we will need to deal with real-time detection, mitigation and possibly attack elimination, as well as data collection for deferred ulterior analysis.

\subsection{Modelling}

Research in communication systems has a solid background of modelling methodologies such as stochastic processes, queueing systems, graph models, etc. These methods are routinely and successfully utilised to describe communication systems and to analyse and improve their performance, but they are rarely used for security. Our research plan for anomaly detection incorporates modelling of the wireless communication network at different levels of abstraction to properly represent the components and the processes that are prone to anomalous behaviour.

The natural choice for this approach is multi-class queueing and diffusion models \cite{Acta,Labed} and related methods. Such models provide estimates of both averages and variances of the times that it would take to undertake signaling or call processing functions, as well as of access times to web sites and other services, in the presence of a large population of mobile users in the network, once the average service times and task sequences are known. For a given (small) subset of users which one wishes to monitor, if the estimated average quantities for a population scaled up to the current observed numbers in the network, for the same user set, deviates significantly from the current measured values in the network, then one can infer some level of anomaly. The estimates for the scaled up population can be calculated from the queueing models in a very fast and straightforward manner, leading to a useful modelling based anomaly detector. This performance based approach can also provide billing estimates based on the utilisation and durations related to internal and external resources, so that the queueing model results can map into billing information: once again, deviations from the expected values (even in the presence of large traffic loads) can be estimated by comparing model predictions with measured observations.

In addition to the above approach, some of the analysis may require more effort because contrary to standard approaches that obtain the steady-state behaviour, the detection of anomalies may require detection of change that is time dependent and thus requires transient or rare event analysis. In order to address the rise of the resulting computational complexity, we also plan to utilise the learning capabilities of the Random Neural Network (RNN) \cite{Gelenbe89,Gelenbe93,Multiple,Gelenbe2002,Gelenbe2010,Natural}, which has previously been applied to a variety of complex problems. The RNN's mathematical analogy to a queueing and communication network will ease the design of learning techniques that have to mimic the behaviour of mobile calls in a wireless network with its different resources and customers.

\subsection{Simulation Tools}

Our analytical models and anomaly detection algorithms will be augmented and validated with simulation tools. As an initial step, we are developing realistic simulations of UMTS and LTE networks using the OPNET simulator in order to extract data regarding control-plane events that take place during normal mobile communications. For this purpose, we are currently modelling a small-scale mobile network and all control-plane protocols in the packet-switched domain, including mobility, session and radio resource management. Characteristics of these control events will be used to drive the development of our analytical models. We will later increase the scale of our simulations to validate our mathematical results. For performance reasons in this stage, instead of simulating communications and events of all mobile users, we will identify the generic characteristics of a large number of users and use these to generate background traffic on the network while explicitly simulating a smaller set of users, among which only a few may demonstrate anomalous behaviour. Another set of simulations will include billing system components to monitor monetary use of internal and external network resources, and based on data traces and parameters obtained from real mobile networks, these simulations will be used to generate synthetic data for the learning methods we plan to apply and to test the performance of our real-time and offline anomaly detection methods. Finally, we will employ simulation as a tool for the integration and validation of other system components which are developed by NEMESYS partners, such as the attack correlation and visualisation \& analysis modules~\cite{Tzovaras2013}.

\section{Prior Work on Network Threats and Mitigation}

Mobile networks are vulnerable to a form of denial-of-service (DoS) attack known as \emph{signaling attacks} in which the control plane is overloaded through low-rate traffic patterns that exploit vulnerabilities in the signaling procedures involved in paging \cite{Serror2006}, transport of SMS \cite{Enck2005} and radio resource control \cite{Lee2009} (see \cite{Ricciato2010} for a review on the subject). In principle, signaling attacks can be carried out either by compromising a large number of mobile devices as in the case of distributed DoS attacks on wired networks \cite{Gelenbe2007} or from outside the network (e.g. the Internet) by targeting a hit list of mobile devices through carefully timed traffic bursts. In order to orchestrate such attacks, active probing can be used to infer the radio resource allocation policies of operational networks \cite{Barbuzzi2008,Qian2010} as well as to identify a sufficient number of IP addresses in a particular location \cite{Qian2012}. Moreover, a study \cite{Wang2011,Qian2012} of 180 cellular carriers around the world revealed that 51\% of them allow mobile devices to be probed from the Internet. Despite their feasibility, signaling attacks are yet to be observed in practice, which is likely due to the lack of financial incentives for cyber criminals who would rather have the infrastructure functional in order to launch profitable attacks. A related threat known as \emph{signaling storms} occur often as a result of poorly designed popular mobile apps that repeatedly establish and tear down data connections, generating huge levels of signaling traffic capable of crashing a mobile network. Thus signaling storms have the same effect as a DoS attack \cite{Gelenbe2007,CACM-SAN}, but without the malicious intent, and they are becoming a serious threat to the availability and security of cellular networks. For example, a Japanese mobile operator suffered a major outage in 2012 \cite{Storm2012}, which was attributed to an Android VoIP app that constantly polls the network even when users are inactive. Moreover, according to a recent survey of mobile carriers \cite{Arbor2012}, many of them have reported outages or performance issues caused by non-malicious but misbehaving apps, yet the majority of those affected followed a reactive approach to identify and mitigate the problem. Note that unlike flash crowds, which normally happen and last for a short period of time coinciding with special occasions such as New Year's Eve, signaling storms are unpredictable and tend to persist until the underlying problem is identified and corrected. This has prompted the mobile industry to promote best practices for developing network-friendly apps \cite{GSMA2012,Jiantao2012}. However, the threat posed by large scale mobile botnets cannot be eliminated in this manner, as botmasters care more about increasing their revenue and stealthiness than the impact that their activities have on the resources of mobile networks.

\noindent \emph{Countermeasures:} signaling problems have a limited impact on the data plane and thus are difficult to detect using traditional intrusion detection systems which are effective against flooding type attacks. For Internet-based attacks, a change detection algorithm using the cumulative sum method has been proposed in \cite{Lee2009}, where the signaling rate of each remote host is monitored and an alarm is triggered if this rate exceeds a fixed threshold. The use of a single threshold for all users, however, presents a trade-off between false positives and detection time, which can be difficult to optimise given the diversity of users' behaviour and consumption. A supervised learning approach is used in \cite{Gupta2013} to detect mobile-initiated attacks whereby transmissions that trigger a radio access bearer setup procedure are monitored, and various features are extracted relating to destination IP and port numbers, packet size, variance of inter-arrival time, and response-request ratio. One problem with supervised learning techniques is that both normal and malicious behaviours need to be defined in advance, rendering them ineffective against new and sophisticated types of attacks. Detection of SMS flooding attacks is considered in \cite{Kim2013}, where low reply rate is used as the main indicator of malicious activities, which is likely to misclassify SMS accounts used for machine-to-machine (M2M) communications, such as asset tracking and smart grid meters \cite{Murynets2012}.

\subsection{Attacks Against Mobile Users}

A recent report by Kaspersky \cite{Kaspersky2013} revealed that the most frequently detected malware threats affecting Android OS are SMS trojans, adwares and root exploits. Mobile botnets are also emerging as a serious threat because of their dynamic nature, i.e. they could be used to execute any action at the command of a botmaster. In the following we summarise the various approaches that have been proposed to enable mobile operators to detect attacks against users.

\emph{Network level analysis} has been explored in a number of recent studies focusing on three aspects:
\begin{itemize}[leftmargin=*,topsep=0.2em]
	\item Domain Name System (DNS): Since malware typically uses DNS to retrieve IP addresses of servers, detecting and blacklisting suspicious domains can be a first step towards limiting the impact of malware \cite{Iland2012,Lever2013}. However, detection should not be based solely on historical data (i.e. known malicious domains), but also on behavioural characteristics such as host changes and growth patterns which may differentiate normal and abnormal traffic.

	\item Call Charging Records (CDR): One of the key characteristics of mobile communications pertains to the fact that the whole extent of exchanged traffic load is continuously monitored for billing and accounting purposes. Hence, it is expected that many malicious activities will have an evident impact on the CDR of the parties involved. In \cite{Murynets2012}, communication patterns of SMS spammers are compared to those of legitimate mobile users and M2M connected appliances, showing evidence of spammer mobility, voice and data traffic resembling the behaviour of normal users, as well as similarities between spammers and M2M communication profiles. Fuzzy-logic is used in \cite{Vural2010} to detect SMS spamming botnets by exploiting differences in usage profiles, while in \cite{Yan2009} SMS anomalies are detected through building normal social behaviour profiles for users, but the learning technique fails to identify transient accounts used only for malicious purposes. Markov clustering of international voice calls \cite{Jiang2012} indicates that different fraud activities, such as those carried by malicious apps with automated dialler or via social engineering tactics, exhibit distinct characteristics.

	\item Content matching: Uncommon header flags and syntactic matches in HTTP messages can be used as indicators of data exfiltration attempts \cite{Iland2012}, but this approach is not effective when end-to-end encryption is used, as it relies on extracting information from plain-text transmissions.

\end{itemize}

\emph{Cloud-based detection} offers a trade-off between network level analysis and on-device security: the former imposes zero-load on the device, but limits its scope to cellular data, while the latter is able to utilise internal mobile events for detection but is resource hungry. There are two main approaches, both of which offload intensive security analysis and computations to the cloud. The first uses a thin mobile client to extract relevant features from the device \cite{Schmidt2009} including free memory, user activity, running processes, CPU usage and sent SMS count, which are then sent to a remote server for inspection. An example is Crowdroid \cite{Burguera2011} which collects system calls of running apps and sends them preprocessed to a server that applies a clustering algorithm to differentiate between benign and trojanised apps. Although this approach can offer heavy-weight security mechanisms to devices that may not otherwise have the processing power to run them, it still requires continuous monitoring, some processing and frequent communication with a cloud service, thus limiting its utility. In the second approach \cite{Portokalidis2010,Houmansadr2011,Zhao2012} an exact replica of the mobile device is stored in a virtual environment in the cloud, and the system consists of three modules: (i) a {\em server} applying heavy-weight security analyses on the replica, such as virus scanning \cite{Portokalidis2010,Zhao2012} and off-the-shelf intrusion detection systems \cite{Portokalidis2010,Houmansadr2011}; (ii) a {\em proxy} duplicating incoming traffic to the mobile device, and forwarding it to the mirror server; and (iii) a {\em tracer} on the mobile recording and offloading all necessary information needed to replay execution on the replica. The advantage of this approach is that it can leverage existing complex security solutions, but the processing and energy costs of synchronising the entire state of a device are prohibitive.

\section{Conclusions}

The goal of the NEMESYS project is to develop a novel security framework for gathering and analysing information about the nature of cyber-attacks targeting mobile devices and networks, and to identify abnormal events and malicious network activity \cite{NEMESYS1}. Thus this paper summarises our proposed approaches to the analysis of network traffic and the development of anomaly detection algorithms, combining modelling and learning using network measurements, control-plane and billing data. Since network threats often map into \emph{congestion} in the signaling system, we will use queueing models to understand bottlenecks in signaling protocols to identify and predict abnormal traffic patterns that arise in such phenomena. In addition, we will develop efficient algorithms to detect and classify attacks based on semi-supervised and unsupervised learning, that can process massive amounts of signaling and billing data in real-time, allowing the early warning of abnormal activities. Finally, we will use the OPNET simulator to run tests and identify possible issues, to contribute to the adjustment of network operational parameters and help mitigate such threats.


\begin{thebibliography}{10}
\providecommand{\url}[1]{{#1}}
\providecommand{\urlprefix}{URL }
\expandafter\ifx\csname urlstyle\endcsname\relax
  \providecommand{\doi}[1]{DOI~\discretionary{}{}{}#1}\else
  \providecommand{\doi}{DOI~\discretionary{}{}{}\begingroup
  \urlstyle{rm}\Url}\fi

\bibitem{Aguilar}
Aguilar, J., Gelenbe, E.: Task assignment and transaction clustering heuristics
  for distributed systems.
\newblock Information Sciences \textbf{97}(1-2), 199--219 (1997)

\bibitem{Arbor2012}
{Arbor Networks}:
  \href{http://www.arbornetworks.com/research/infrastructure-security-report}{Worldwide
  Infrastructure Security Report} (2012)

\bibitem{AVG2011}
{AVG}:
  \href{http://mediacenter.avg.com/content/mediacenter/en/news/new-avg-study.html}{New
  {AVG} Study Reveals Smartphone Users Not Aware of Significant Mobile Security
  Risks} (2011)

\bibitem{Barbuzzi2008}
Barbuzzi, A., Ricciato, F., Boggia, G.: Discovering parameter setting in {3G}
  networks via active measurements.
\newblock IEEE Commun. Lett. \textbf{12}(10), 730--732 (2008)

\bibitem{BBC2012}
{BBC}: \href{http://www.bbc.co.uk/news/technology-20080397}{Cyber thieves
  profit via the mobile in your pocket} (2012)

\bibitem{Burguera2011}
Burguera, I., Zurutuza, U., Nadjm-Tehrani, S.: Crowdroid: behavior-based
  malware detection system for android.
\newblock In: Proc. SPSM '11, pp. 15--26. ACM, Chicago, IL (2011)

\bibitem{Canalys2012}
Canalys:
  \href{http://www.canalys.com/newsroom/smart-phones-overtake-client-pcs-2011}{Smart
  phones overtake client {PCs} in 2011} (2012)

\bibitem{Cisco2013}
Cisco:
  \href{http://www.cisco.com/en/US/solutions/collateral/ns341/ns525/ns537/ns705/ns827/white_paper_c11-520862.pdf}{{Cisco}
  visual networking index: Global mobile data traffic forecast update,
  2012--2017}.
\newblock White Paper (2013)

\bibitem{Enck2005}
Enck, W., et~al.: Exploiting open functionality in {SMS}-capable cellular
  networks.
\newblock In: Proc. CCS '05, pp. 393--404. ACM, Alexandria, VA (2005)

\bibitem{Acta}
Gelenbe, E.: Probabilistic models of computer systems.
\newblock Acta Inform. \textbf{12}(4), 285--303 (1979)

\bibitem{Gelenbe89}
Gelenbe, E.: Random neural networks with negative and positive signals and
  product form solution.
\newblock Neural Comput. \textbf{1}(4), 502--510 (1989)

\bibitem{Gelenbe93}
Gelenbe, E.: Learning in the recurrent random neural network.
\newblock Neural Comput. \textbf{5}, 154--164 (1993)

\bibitem{CACM-SAN}
Gelenbe, E.: Steps towards self-aware networks.
\newblock Commun. ACM \textbf{52}(7), 66--75 (2009)

\bibitem{Natural}
Gelenbe, E.: Natural computation.
\newblock Comput. J. \textbf{55}(7), 848--851 (2012)

\bibitem{Multiple}
Gelenbe, E., Fourneau, J.M.: Random neural networks with multiple classes of
  signals.
\newblock Neural Comput. \textbf{11}(4), 953--963 (1999)

\bibitem{Gelenbe2002}
Gelenbe, E., Hussain, K.: Learning in the multiple class random neural network.
\newblock IEEE Trans. Neural Netw. \textbf{13}(6), 1257--1267 (2002)

\bibitem{Labed}
Gelenbe, E., Labed, A.: G-networks with multiple classes of signals and
  positive customers.
\newblock Eur. J. Oper. Res. \textbf{108}(2), 293--305 (1998)

\bibitem{Gelenbe2007}
Gelenbe, E., Loukas, G.: A self-aware approach to denial of service defence.
\newblock Comput. Netw. \textbf{51}(5), 1299--1314 (2007)

\bibitem{Muntz}
Gelenbe, E., Muntz, R.R.: Probabilistic models of computer systems: Part i
  (exact results).
\newblock Acta Inform. \textbf{7}(1), 35--60 (1976)

\bibitem{Gelenbe2010}
Gelenbe, E., Timotheou, S., Nicholson, D.: Fast distributed near-optimum
  assignment of assets to tasks.
\newblock Comput. J. \textbf{53}(9), 1360--1369 (2010)

\bibitem{NEMESYS1}
Gelenbe, E., et~al.: {NEMESYS}: Enhanced network security for seamless service
  provisioning in the smart mobile ecosystem.
\newblock In: Proc. ISCIS 2013, LNEE. Springer-Verlag (2013)

\bibitem{GSMA2012}
GSMA:
  \href{http://www.gsma.com/technicalprojects/wp-content/uploads/2012/04/gsmasmarterappsforsmarterphones0112v.0.14.pdf}{Smarter
  Apps for Smarter Phones!} (2012)

\bibitem{Gupta2013}
Gupta, A., et~al.: Detecting {MS} initiated signaling {DDoS} attacks in {3G/4G}
  wireless networks.
\newblock In: Proc. COMSNETS'13, pp. 1--6 (2013)

\bibitem{Houmansadr2011}
Houmansadr, A., Zonouz, S.A., Berthier, R.: A cloud-based intrusion detection
  and response system for mobile phones.
\newblock In: Proc. DSNW '11, pp. 31--32. IEEE Computer Society, Hong Kong
  (2011)

\bibitem{Iland2012}
Iland, D., Pucher, A., Sch\"{a}uble, T.: Detecting android malware on network
  level.
\newblock Tech. rep., UC Santa Barbara (2012)

\bibitem{Jiang2012}
Jiang, N., et~al.: Isolating and analyzing fraud activities in a large cellular
  network via voice call graph analysis.
\newblock In: Proc. MobiSys '12, pp. 253--266. ACM, Lake District, UK (2012)

\bibitem{Jiantao2012}
Jiantao, S.: Analyzing the network friendliness of mobile applications.
\newblock Tech. rep., Huawei (2012)

\bibitem{Kim2013}
Kim, E.K., McDaniel, P., Porta, T.: A detection mechanism for {SMS} flooding
  attacks in cellular networks.
\newblock In: SecureComm'12, \emph{LNICST}, vol. 106, pp. 76--93. Springer
  (2013)

\bibitem{Lee2009}
Lee, P.P.C., Bu, T., Woo, T.: On the detection of signaling {DoS} attacks on
  {3G/WiMax} wireless networks.
\newblock Comput. Netw. \textbf{53}(15), 2601--2616 (2009)

\bibitem{Lever2013}
Lever, C., et~al.: The core of the matter: Analyzing malicious traffic in
  cellular carriers.
\newblock In: Proc. NDSS'13, pp. 1--16. San Diego, CA (2013)

\bibitem{lookout2013}
{Lookout Mobile Security}:
  \href{https://blog.lookout.com/blog/2013/04/19/the-bearer-of-badnews-malware-google-play}{The
  Bearer of {BadNews}} (2013)

\bibitem{Kaspersky2013}
Maslennikov, D.:
  \href{http://www.securelist.com/en/analysis/204792283/Mobile_Malware_Evolution_Part_6}{Mobile
  Malware Evolution: Part 6}.
\newblock Tech. rep., Kaspersky Lab (2013)

\bibitem{Murynets2012}
Murynets, I., Jover, R.P.: Crime scene investigation: {SMS} spam data analysis.
\newblock In: Proc. IMC '12, pp. 441--452. ACM, Boston, MA (2012)

\bibitem{Tzovaras2013}
Papadopoulos, S., Tzovaras, D.: Towards visualizing mobile network data.
\newblock In: Proc. ISCIS 2013. Springer-Verlag (2013)

\bibitem{Portokalidis2010}
Portokalidis, G., et~al.: Paranoid {Android}: versatile protection for
  smartphones.
\newblock In: Proc. ACSAC '10, pp. 347--356. ACM, Austin, TX (2010)

\bibitem{Qian2010}
Qian, F., et~al.: Characterizing radio resource allocation for {3G} networks.
\newblock In: Proc. IMC '10, pp. 137--150. ACM, Melbourne, Australia (2010)

\bibitem{Qian2012}
Qian, Z., et~al.: You can run, but you can't hide: Exposing network location
  for targeted {DoS} attacks in cellular networks.
\newblock In: Proc. NDSS'12, pp. 1--16. San Diego, CA (2012)

\bibitem{Storm2012}
{Rethink Wireless}:
  \href{http://www.rethink-wireless.com/2012/01/30/docomo-demands-googles-signalling-storm.htm}{{DoCoMo}
  demands {G}oogle's help with signalling storm} (2012)

\bibitem{Ricciato2010}
Ricciato, F., Coluccia, A., D'Alconzo, A.: A review of {DoS} attack models for
  {3G} cellular networks from a system-design perspective.
\newblock Comput. Commun. \textbf{33}(5), 551--558 (2010)

\bibitem{Schmidt2009}
Schmidt, A.D., et~al.: Monitoring smartphones for anomaly detection.
\newblock Mobile Netw. Appl. \textbf{14}(1), 92--106 (2009)

\bibitem{Serror2006}
Serror, J., Zang, H., Bolot, J.C.: Impact of paging channel overloads or
  attacks on a cellular network.
\newblock In: Proc. WiSe '06, pp. 75--84. ACM, Los Angeles, CA (2006)

\bibitem{Vural2010}
Vural, I., Venter, H.: Mobile botnet detection using network forensics.
\newblock In: Proc. FIS'10, \emph{LNCS}, vol. 6369, pp. 57--67. Springer-Verlag
  (2010)

\bibitem{Wang2011}
Wang, Z., et~al.: An untold story of middleboxes in cellular networks.
\newblock In: Proc. SIGCOMM 2011, pp. 374--385. ACM, Toronto, Canada (2011)

\bibitem{Yan2009}
Yan, G., Eidenbenz, S., Galli, E.: {SMS}-watchdog: Profiling social behaviors
  of {SMS} users for anomaly detection.
\newblock In: Proc. RAID '09, pp. 202--223. Springer-Verlag, Saint-Malo, France
  (2009)

\bibitem{Zhao2012}
Zhao, B., et~al.: Mirroring smartphones for good: A feasibility study.
\newblock In: MobiQuitous'10, \emph{LNICST}, vol.~73, pp. 26--38. Springer
  (2012)

\end{thebibliography}

\end{document}